\documentclass[preprint,showpacs,preprintnumbers,amsmath,amssymb]{revtex4}

\def\aits{{A{\"\i}t-Sahalia's }}

\usepackage{amsmath}

\usepackage{graphicx}
\usepackage{dcolumn}
\usepackage{bm}

\begin{document}

\preprint{APS/123-QED}

\title{On the Use of Local Diffusion Models for Path Ensemble Averaging in Potential of Mean Force Computations }

\author{Christopher P. Calderon}%
 \email{ccaldero@princeton.edu}
\affiliation{%
Department of Chemical Engineering, Princeton University,
Princeton, New Jersey 08544-5263, USA
}%

\date{\today ; First Draft: March 17, 2006}

\begin{abstract}
We use a constant velocity steered molecular dynamics (SMD)
simulation of the stretching of deca-alanine in vacuum to
demonstrate a technique that can be used to create surrogate
stochastic processes using the time series that come out of SMD
simulations. The surrogate processes are constructed by first
estimating a sequence of local parametric models along a SMD trajectory  
and then a single global model is constructed by piecing the local
models together through smoothing splines (estimation is made
computationally feasible by likelihood function approximations).
 The calibrated surrogate models are then
``bootstrapped" in order to simulate the large number of work
paths typically needed to construct a potential of mean force
(PMF) by appealing to Jarzynski's work theorem. 
 When this procedure is repeated for a small
number of SMD paths, it is shown that the global models appear to
come from a single family of closely related diffusion processes.
Possible techniques for exploiting this observation are also
briefly discussed. The findings of this paper have potential
relevance to computationally expensive
 computer simulations and
    experimental works involving optical tweezers where
    it difficult to collect a large number
    of samples, but possible to sample accurately and frequently in time. 
\end{abstract}

\maketitle

\small
\section{\label{intro} Introduction}
The truncated Taylor series of a nonlinear function is probably
the most widely known example of a local model approximation. The
work in \cite{llglassy,lllv} demonstrated some possible extensions
of this basic concept to state-space diffusion models that aimed
at utilizing the desirable features of a finite dimensional
parametric estimator, but at the same time retained the
flexibility associated with a ``Taylor series".  In this paper we
demonstrate how a local diffusion modeling approach can be applied
to time-inhomogeneous, state-dependent noise processes. The goal
is to obtain a  stochastic differential equation (SDE), with
nonlinear coefficient functions, that accurately describes the
dynamics of a \emph{single} trajectory associated with the time
series output of a steered molecular dynamics (SMD) simulation.
The system used to illustrate the approach is the well-studied
 example of using
constant velocity SMD to simulate the unravelling (in vacuum) of
deca-alanine at constant temperature (extensive simulation results
exist for this system because it is a  ``fast-folder" \cite{schultenPMF}).

A common objective of many of these types of SMD simulations is to
derive a single low-dimensional effective diffusion equation valid
at mesoscale times using a batch of time series  of atomistic
simulation output \cite{hummerPNAS01,ioan,schultenJCP04}.  This
type of multiscale modeling approach often requires one to compute
the potential of mean force (PMF) associated with the system.
 One technique for calculating the PMF associated with a  set of well
 selected
 reaction coordinates \cite{schultenPMF}  is to appeal to the Jarzynski work theorem
\cite{jarz97} using the path dependent work (measured
from the SMD simulations) in order to reconstruct the PMF associated
with the selected reaction coordinate(s) of the system.  To do this, one
typically needs to assume variety of things about the system a priori \cite{jarz97,crooks99};  
even if all of the assumptions hold true for the selected reaction
coordinate(s) of the system, calculating the PMF can be difficult
in practice due in part to the number of sample paths needed in
order to accurately approximate the PMF from atomistic simulation
trajectories \cite{schultenJCP04,jarzysnki06}.  A large number of
paths are needed because  realizations  with a low probability of
occurrence can cause a large influence on the PMF (which depends
on an exponential average) computed from a
 finite sample of realizations \cite{jarzysnki06}.
 The requirement
of a large number of atomistic simulation trajectories is
problematic because current computational technology limits the
amount of data that can be generated in a reasonable amount of CPU
time.
  As a result, numerous schemes \cite{hooker_MLE,ioan,schultenJCP04} have
   been developed for efficiently sampling from SMD
   simulations  (the
     cited works focus on making efficient use of
     ensemble data created from a batch of genuine SMD runs).

   The main focus in this work is to approximate and characterize the statistical
   properties of SMD simulations on a pathwise basis and then repeat the procedure over
   a small number of (frequently sampled) genuine SMD time series.
    It should be stressed that we determine a global diffusion process
    using the time series associated with a \emph{single} SMD run; the information
    contained in a batch of SMD processes is not pooled together to
    determine a single diffusion process (each SMD process realization
    results in the estimation of a new diffusion model).   
   If the surrogate process's dynamics
   are close to that of the genuine process  then we have
   a means for generating the  large number of (approximate) work paths
   typically needed in the Jarzynski work relationship.
   The large number of samples needed can come
   from bootstrapping the estimated models
   (that is, estimate the model parameters using actual SMD runs and then
   use the estimated models to give additional samples
    by using different random number sequences).
    The bootstrapped surrogate models used here are
     computationally cheap (relative to the SMD simulations)
      diffusion SDEs.

The purpose of the diffusion models we estimate from SMD data is
\emph{not} to obtain  equations from atomistic data valid at
mesoscopic timescales.  The intention is to approximate the work
distribution associated with SMD simulations using  simple
diffusion models in hopes of constructing a surrogate process
which can be used to replace the computationally expensive SMD
runs. Calibrating a diffusion equation valid at meso- or even
macroscopic timescales \emph{directly} from atomistic data is
problematic because 
the reaction coordinate description of the process neglects
several details of the underlying system.  It becomes difficult to
quantify how the model imperfections and the errors introduced by
finite
sample estimation 
 interact and
ultimately affect the mesoscale equation.  The local diffusion
models we use are motivated by the overdamped limit \cite{zwanzig}
of the Langevin equation. Throughout we operate in a regime that
is intermediate to the under and overdamped limits; we simply find
the ``projection" (through estimation) of the data onto this
particular diffusion model class.  In this paper  we demonstrate
our local  estimation technique by modeling the dynamics of an
established \cite{schultenJCP04} reaction
    coordinate ($z \equiv $ the end-to-end distance
     of the deca-alanine molecule)
     using SMD trajectories that were generated by the NAMD program \cite{namdprog}. In Section \ref{s_res} we demonstrate that we can  approximate the work
distributions fairly accurately (the accuracy is sufficient to closely reproduce the PMF of the system) which gives
partial evidence that the overdamped approximation is reasonable for the
system conditions studied (we later give additional statistical evidence which further justifies our overdamped approximation).

Furthermore the dynamics of the reaction coordinate \emph{alone}
are not likely Markovian (which is implicitly assumed in our
diffusion model) due in part to some of the reasons mentioned in
the previous paragraph.  The Jarzynski work relationship
\cite{crooks99}  requires that all of the degrees of freedom of
the system collectively obey Markovian dynamics (the work
relationship can still be valid if the evolution of the reaction
coordinate alone is non-Markovian, e.g. if the dynamics of the
\emph{full} system come from a Hamiltonian \footnote{These
comments were made after a private communication with C.
Jarzynski.}).  If one could somehow reliably model the dynamics
resulting from the interaction of the ``heat bath" \footnote{The
use of the term ``heat bath" is somewhat nonconventional here.  By
``heat bath" I mean the random effects caused by the surrounding
medium (randomness due to classical heat transfer) and effects due
to not explicitly modeled fast internal degrees of freedom (e.g.
vibrational degrees of freedom).} with the reaction coordinate,
then one would have more faith in appealing to a Markovian
description of the reaction coordinate dynamics.  We merely lump
all of the effects of the heat bath into a Brownian motion (with
state dependence on the local noise magnitude); the hope is that
the surrogate model is good enough to approximate work
distributions, but the simple nature of the model casts doubt on
the validity of extrapolating the estimated models to larger
length and time scales.
The surrogate model we introduce can
potentially be plugged into other schemes which aim at
\emph{indirectly} approximating the overdamped limit equation
\cite{schultenJCP04,ioan} (with the intention of determining a
diffusion that is valid at meso- or even macroscopic time scales),
but we stress (again) that the estimated models presented here should not
be used directly to ambitiously extrapolate to larger length and
time scales.

The use of the probability integral transform (PIT) \cite{diebold}
using the simulated maximum likelihood (SML) approximation of the
transition densities associated with a global model constructed by
piecing together local diffusion models is also investigated. The
PIT is used to create goodness-of-fit test statistics \cite{hong}
associated with the models calibrated from the data of our
nonstationary SMD processes. 
  The PIT allows one to judge the statistical validity of
the assumed stochastic model given data generated by the true
process. The PIT can sometimes even offer insight concerning model
inadequacies \cite{diebold,hong}. In Section \ref{s_res} we give
quantitative evidence which suggests that the errors of the
overdamped approximation we impose on the model are small in
comparison to the errors associated with summarizing the dynamics
of the system with a scalar end-to-end distance reaction
coordinate (using a simple diffusion model).

 A major finding of this work is that in some
systems it is  possible to approximate the work distributions
needed to compute the PMF from a relatively small number  of SMD
trajectories by appealing to estimation along individual sample
paths and using the calibrated surrogate models to generate the
large number of ``synthetic" trajectories needed in standard
applications of constructing a PMF by using the Jarzynski
nonequilibrium work relationship. It is shown (for the particular
system studied) that the estimated global diffusion approximations
associated with the different SMD trajectories result in what
appears to be a single family of closely related nonlinear
diffusion processes. In systems where the reaction coordinate has
more complicated interactions with ``the surroundings" (e.g.
environmental asymmetries
\cite{schultenporecond,kozstinschulten,schultenCYANO_04,hummercarbon}
or jump-like transitions \cite{hummer_science}) there is no
guarantee that the diffusion approximations will belong to a
single family. However, the findings reported here  indicate that
it may be possible to appeal to empirical Bayes \cite{berger} and
growth curve analysis \cite{ramsay} techniques  in order to
approximate the SMD process. This would be computationally
tractable if the SMD process under study can be adequately
characterized by a relatively small number of ``diffusion
families".  This point is demonstrated using a toy model in
Section \ref{s_extension}.
   The findings of this paper have potential relevance to computationally expensive
 computer simulations and
    experimental works where
    it difficult to collect a large number
    of samples, but possible to sample accurately and frequently in time \cite{hummer_science,hummerPNAS01,schultenJCP04}.

The remainder of the paper is organized as follows: Section
\ref{s_localmod} reviews the local modeling approach used. Section
\ref{s_stattools} reviews some basic facts about the statistical
tools that we use for estimation and inference. Section
\ref{s_pmf} gives the equations that we use to estimate a PMF
using our ``synthetically" created data. In Section
\ref{s_simdetails} we briefly report some of the computational
details. Section \ref{s_res} gives our results; Section
\ref{s_extension} contains a discussion and
 presents a toy example which aims at
demonstrating how the findings in this paper may be exploited; we
then conclude.

\section{\label{s_localmod} Local Parametric modeling Approach}
We attempt to fit a global SDE of the generic form:
\begin{equation}
 dX_t=b(X_t,t;\Theta)dt+
 \sigma(X_t;\Theta)
 dW_t.
 \label{SDEgeneric}
\end{equation}
to the output of a single SMD simulation (other independent SMD
runs result in new SDEs) by appealing to estimation techniques
related to maximum likelihood (ML).  In the above equation, $X_t$
corresponds to the reaction coordinate whose dynamics are being
approximated, $W_t$ is the standard Brownian motion,
$b(\cdot,\cdot;\Theta)$ and $\sigma(\cdot;\Theta)$ are the assumed
drift and diffusion coefficient functions (parameterized by
 $\Theta$), and all SDEs correspond to It\^{o} integrals. The process above is  referred to as the ``diffusion process" in the sequel.


One typically does not  know \emph{a priori}  a parametric family
of functions that  the drift and diffusion coefficients belong to
that can adequately  describe the global dynamics (it should be
noted that the $\Theta$ ``parameterization" is not a traditional
Euclidean parameter in our global models). A nonparametric
approach \cite{dietz01,kuto} can sometimes successfully overcome
this difficulty, but
 nonparametric estimators  are typically not as
 efficient (asymptotically) as the finite dimensional parametric ML
estimator. More importantly, a parametric framework  allows us to
impose a smooth structure that ``ignores" the often more
complicated structure associated with fast timescales. To do this
with confidence, care must be taken \cite{stuart06}; we present
methods that can quantitatively assist one in this type of
modeling.

 In this paper, we propose a \emph{local} finite
dimensional parametric estimator based on linear functions (these
functions are then used to construct the drift and the diffusion
coefficients).  The nature of the time-dependent forcing term is
typically prespecified in SMD simulations \cite{schultenJCP04}
which greatly facilitates selecting the functional form of  our
local models . In the constant velocity SMD application studied,
our local models take the form:

\begin{align}
\nonumber dX_t= & b^{\mathrm{LOC}}(X_t,t;\theta)dt+ \sigma^{\mathrm{LOC}}(X_t;\theta) dW_t\\
\nonumber b^{\mathrm{LOC}}(X,t;\theta)\equiv & \beta \frac{\sigma^{\mathrm{LOC}}(X;\theta)^2}{2}\mu^{\mathrm{LOC}}(X,t;\theta) \\
\nonumber \sigma^{\mathrm{LOC}}(X;\theta)\equiv& \sqrt{2} \Big(C+D(X-X_o)\Big) \\
\nonumber \mu^{\mathrm{LOC}}(X,t;\theta)\equiv &\Big(A+B(X-X_o)\Big) + k_{\mathrm{pull}}\Big(X^{targ}(t)-X\Big). \\
\end{align}

In the above, $X_o$ is a specified point at which the  local model
is centered and the parameter vector estimated is $\theta \equiv
(A,B,C,D) $; the elements of this vector correspond to the
constants used for a local approximation of the   nonlinear global
(assumed deterministic but unknown) functions $\sigma(\cdot)$ and
$b(\cdot,\cdot)$; $k_{\mathrm{pull}}$ is the harmonic constraint
used in the constant velocity SMD simulation and $X^{targ}(t)$ is
a deterministic function (in this study,
$X^{\mathrm{targ}}(t):=X^{\mathrm{IC}}+v_{\mathrm{pull}}t$, where
$X_{IC}$ and $v_{\mathrm{pull}}$ are specified constants). The
constant $\frac{1}{\beta} \equiv {k_BT}$ was set to 0.6 in reduced
units (using 1 kcal/mol as the energy scale) where  $k_B$
corresponds to Boltzmann's constant and $T$ is the  system
temperature (300 K maintained by a Langevin heat bath). These
local models are estimated at points ($X_o$) selected in an ad hoc
manor (an optimal strategy for picking these points should be
investigated in the future) and a global model is obtained by
connecting the estimated constants $A$ and $C$ at the various
$X_o$'s selected using smoothing splines \cite{deboor} (MATLAB's
\url{csaps} function was used for spline smoothing).  It should be
noted that one could also develop interpolation schemes that
utilize the information contained in the estimated parameters $B$
and $D$ (e.g. see \cite{llglassy}); this was not done in order to
keep the presentation streamlined. It should also be noted that
although we deal exclusively with the scalar case, the
 techniques presented here can trivially be extended to the case
 of multivariate time-inhomogeneous, state-dependent noise systems
 (which would be relevant to SMD experiments which require both the
 end-to-end separation of a molecule as well as some dihedral
 angle(s) \cite{yanniscoarsedihedral} to adequately characterize the system). The
 major
 computational obstacle in the
 multivariate case is the  number of parameters to be estimated
 (which
 can grow rapidly
 as a function of the state dimension).

The most questionable assumption of the technique presented here
is that the dynamics of the reaction coordinate associated with
different SMD trajectories can be well-represented by simple
diffusion processes. One potential way of incorporating a more
realistic noise process is briefly discussed in Section
\ref{s_extension}. In addition, the local process approximation
technique could in principle be extended to jump-diffusion
processes \cite{singleton01}, but many fundamental estimation
issues still need to be resolved before one attempts to use local
models to construct global time-inhomogeneous jump-diffusion
models that can be used to reliably approximate SMD trajectories.

Unfortunately, even for our overly simple local diffusion models,
an analytic expression for the transition density needed for a ML
estimator of the local model is not even available in closed-form
(when a complicated function like a spline is used for the
coefficient functions of the global SDE, an analytic expressions
seems hopeless to obtain). To overcome these difficulties, we
appeal to the SML estimator \cite{sml,smle} in order to carry out
estimation and inference (reviewed in Section \ref{s_stattools}).
This
 method only gives a noisy approximation of the transition density needed for a
 ML type
estimator; other approximations are possible (e.g. a deterministic
extension of \aits expansion \cite{egorov}). One appealing feature
of a deterministic method is that the computation time needed to
find the optimal parameter vector associated with a time series is
typically much less than that associated with simulation based
methods. The parametric estimator we propose does not satisfy all
of the conditions needed to ensure convergence of the
time-dependent Hermite expansion given in \cite{egorov}.
Specifically, the proposed linear diffusion coefficient can take a
value of zero which can cause problems in the Hermite expansions
\cite{egorov,ait2,aitVEC}. Even for models that do satisfy all of
the conditions, a  truncated expansion can yield improper density
expansions which causes additional difficulty in estimation.
Furthermore, one would probably not be able to adapt the method in
\cite{egorov} to account for the spline coefficients we use for
our global model whereas the SML method readily allows one to use
these types of coefficient functions in the SDE model.

When one uses a spline for the global functions in the SML, one
does violate some of the assumptions presented in \cite{sml} that
guarantee convergence of the SML approximation. Specifically
infinite differentiability of the coefficient functions is not
typically satisfied when one uses standard splines, but as
mentioned in \cite{sml} the infinite differentiability assumptions
is probably not necessary to guarantee convergence of the
approximation (coincidentally these assumptions are not violated
in the local parametric  models proposed).

\section{\label{s_stattools} Statistical Tools}
\subsection{Maximum Likelihood Estimation}
We now recall a few basic facts about ML estimation; some standard
references include \cite{hamilton,jeganathan,vandervaart,basawa}.
It is assumed throughout that the \emph{exact} distribution
associated with the parametric model admits a continuous density
whose logarithm is well defined almost everywhere and is at least
three times continuously differentiable with respect to the
parameters \cite{kl}.
ML estimation is based on maximizing the log-likelihood
($\mathcal{L}_{\theta}$) with respect to the parameter vector ($\theta$):
\begin{equation}
\mathcal{L}_{\theta} \equiv \log\Big(f(\mathbf{x};\theta)\Big).
\label{eq_infoint}
\end{equation}

In the above equation, $\mathbf{x}$ corresponds to a matrix of
observations $\in \mathbb{R}^{d\times M}$ where $d$ is the
dimension of the state and $M$ is the length of the time series;
$f(\mathbf{x};\theta)$ corresponds to the probability density
associated with  observation $\mathbf{x}$.
For a single sample path of a discretely observed diffusion known
to be initialized at $\mathbf{x}_0$, $f(\mathbf{x};\theta)$ can be
evaluated as \cite{hamilton}:

\begin{equation}
f(\mathbf{x};\theta)=\delta_{\mathbf{{x}_0}}\prod \limits_{m=1}^{M-1}
f(\mathbf{x}_{m}|\mathbf{x}_{m-1};\theta).
\end{equation}

In this equation $f(\mathbf{x}_{m}|\mathbf{x}_{m-1};\theta)$
represents the conditional probability density (transition
density) of $\mathbf{x}_{m}$ given the observation
$\mathbf{x}_{m-1}$  and $\delta_{\mathbf{{x}_0}}$ is the Dirac
distribution. The associated log-likelihood (given $\theta$, the
data, and the transition density) takes the form:
\begin{equation}
\mathcal{L}_{\theta}:=\sum\limits_{m=1}^{M}
\log\Big(f(\mathbf{x}_{m}|\mathbf{x}_{m-1};\theta)\Big).
\label{eq_loglikelihoodfuncdef}
\end{equation}
We  assume the existence of an invertible symmetric positive
definite ``scaling matrix" matrix $\mathcal{F}_{(M,\ \theta)}$
\cite{lecam00} associated with the estimator; the subscripts are
used to make the dependence of the scaling matrix on $M$ and
$\theta$ explicit.  Under some additional regularity assumptions
\cite{jeganathan,vandervaart}, one has the following limit for a
{\it correctly specified} finite dimensional parametric model:
\begin{equation}
\mathcal{F}_{(M,\
\tilde{\theta})}^{\frac{1}{2}}(\theta_M-\tilde{\theta})
\stackrel{\mathbb{P}_{\tilde{\theta}}}{\Longrightarrow}
\mathcal{N}(\mathbf{0},\mathbf{I}).
\end{equation}

Here $\tilde{\theta}$ is the true parameter vector; $\theta_M$
represents the parameters estimated with a finite time series of
length $M$;
$\stackrel{\mathbb{P}_{\tilde{\theta}}}{\Longrightarrow}$ denotes
convergence in distribution \cite{vandervaart,hamilton} under
$\mathbb{P}_{\tilde{\theta}}$ where the aforementioned
distribution represents that associated with the density
 $f(\mathbf{x};\tilde{\theta)}$;
$\mathcal{N}(\mathbf{0},\mathbf{I})$ denotes a normal distribution
with mean zero and an identity matrix for the covariance.
For a correctly specified model family, $\mathcal{F}_{(M,\
\tilde{\theta})}$ can be estimated (numerically) in a variety of
ways \cite{white,lecam00}.
The appeal of ML estimation lies in that, asymptotically in $M$, the
variance of the estimated parameters
 is the smallest that can be achieved by an
estimator that satisfies the assumed regularity conditions
\cite{jeganathan,vandervaart}. Our motivation for using a sequence
of simple parametric local models comes from this asymptotic
efficiency.

Unfortunately exact analytic expressions for how $\mathcal{F}_{(M,\
\tilde{\theta})}$
scales asymptotically with $M$ are typically difficult to
determine for general nonstationary time series models. Throughout we simply
assume that the scaling matrix associated with the local
models proposed using the approximated transition density is
close to the efficiency associated with the ML estimator.

\subsection{The SML Expansion}

Transition density expansions have been an active area of research
in recent years.  A variety of methods have been introduced that
aim at overcoming the fact that the transition density associated
with a general parametric diffusion model is typically unavailable
in closed-form
\cite{ait2,aitECO,aitextension,bibby,gallant,smle,sml}.  It is
well known \cite{lo} that the simple intuitive ``Euler estimator"
(motivated by the transition density associated with  the
 Euler-Maruyama scheme \cite{kp}) yields significant
bias in parameter estimates.  The Euler-Maruyama discretization
(using a time step of size $\Delta t$) corresponding to our local SDE  is given by:

\begin{align}
 \nonumber X_{t_{m}}= & X_{t_{m-1}}+ b(X_{t_{m-1}},t_{m-1};\theta)\Delta t \\
 & + \sigma(X_{t_{m-1};}\theta)
 \eta_{t_{m-1}}\sqrt{\Delta t}.
 \label{SDEeuler}
\end{align}

Where $m=1\ldots,M$, $X_0$ is the exactly known initial condition,
and $\{\eta_{t_i}\}_{i=0,M-1}$ is a sequence of standard normal
 random variables.  The transition density (given below) associated
with one step of the scheme above is what we refer to as the Euler
estimator; the random variable  $X_{t_m}$ conditional on the value
of $X_{t_{m-1}}$ is denoted by $X_{t_m}|X_{t_{m-1}}$ and the Euler
approximation of its density is given by:

\begin{equation}
X_{t_m}|X_{t_{m-1}} \sim \mathcal{N}\Big(
X_{t_{m-1}} +b(X_{t_{m-1}},t_{m-1};\theta)\Delta t,\sigma(X_{t_{m-1}};\theta)^2 \Delta t \Big).
\end{equation}

The symbol $\sim$ denotes that the random variable on the left of
the symbol is distributed with the law to the right;
$\mathcal{N}(\mu,\sigma^2)$ corresponds to a  normal law with mean
$\mu$ and variance $\sigma^2$ (so the transition density
can be readily evaluated analytically). 
The Euler-Maruyama discretization does not correspond exactly to
the assumed parametric SDE model and the difference can cause
significant problems in estimation \cite{lo,bibby}.  In
\cite{smle,sml} it is proved that an extension of the Euler
estimator idea can yield a fairly reliable estimator. The SML
approximation of the transition density associated with the
observation pair $(X_{t_m},X_{t_{m-1}})$ is obtained by specifying
two additional parameters $T_{\mathrm{sml}}$ and
$N_{\mathrm{sml}}$; these parameters are used to define the
simulation parameter $\delta t \equiv \frac{\Delta
t}{T_{\mathrm{sml}}}$ and

\begin{align}
 \nonumber  X_{\tau_{m^{\prime}}}^{\mathrm{n}}\equiv &
 X_{\tau_{m^{\prime}-1}}^{\mathrm{n}}+  b(X_{\tau_{m^{\prime}-1}}^{\mathrm{n}},\tau_{m^{\prime}-1};\theta) \delta t \\
 & + \sigma(X_{\tau_{m^{\prime}-1}}^{\mathrm{n}};\theta)
 \eta_{\tau_{m^{\prime}-1}}\sqrt{\delta t}.
 \label{SDEeulerSML}
\end{align}


Where $m^{\prime}=1\ldots,T_{\mathrm{sml}}-1$ and $\mathrm{n}=1\ldots,N_{\mathrm{sml}}$; 
in what follows the actually observed transition pair is separated
by a time of length $\Delta t$, but the time indices used by the
SML trajectories are given by $\tau_{m^{\prime}}:=t_{m-1}+
m^{\prime} \delta t$. The distribution associated with each SML
simulation path is given by:
\begin{equation}
\Big(X_{t_m}|X_{t_{m-1}}\Big)^{\mathrm{n}} \sim  \mathcal{N} \Big(
X_{\tau_{T_{\mathrm{sml}}-1}}^{\mathrm{n}}+
b(X_{\tau_{T_{\mathrm{sml}}-1}}^{\mathrm{n}},\tau_{T_{\mathrm{sml}-1}};\theta)\delta
t,\sigma(X_{\tau_{T_{\mathrm{sml}}-1}}^{\mathrm{n}};\theta)^2
\delta t \Big).
\end{equation}

 The transition density above is evaluated for a given $(X_{t_m},X_{t_{m-1}})$ using
  $N_{\mathrm{sml}}$ paths and its value along path
 $\mathrm{n}$  is denoted by
$p_\mathrm{n}(X_{t_m}|X_{t_{m-1}};\theta)$. The transition density
of the observation pair associated with the \emph{original} time
series is obtained by appropriately averaging over the
$N_{\mathrm{sml}}$ paths yielding the following approximation of
the log likelihood function:

\begin{equation}
\mathcal{L}_{\theta}\approx  \sum\limits_{m=1}^{M} \log \Big(
\frac{1}{N_{\mathrm{sml}}}\sum\limits_{n=1}^{N_{\mathrm{sml}}}
p_\mathrm{n}(X_{t_m}|X_{t_{m-1}};\theta)\Big).
\end{equation}

The SML estimator is appealing because it is simple to implement
in a computer program, but for the sample sizes used in SMD
applications its computational cost can be a significant drawback.
The optimal parameter vector using a Nelder-Mead search algorithm
 for a representative  \emph{local} model (repeated for
ten SMD trajectories) took roughly one day  on a PC with a 3.4 GHz
Pentium IV processor which had 1 GB of RAM. The estimation task
associated with multiple trajectories can obviously be trivially
parallelized and the work load associated with this optimization
is much less than that associated with typical SMD simulations,
however a detailed statistical analysis of the data is greatly
hindered (in regards to computation time) by a simulation based
approach such as SML. Other simulation techniques are currently
available, such as Gallant and Tauchen's EMM estimator
\cite{gallant}, however alternative approximations were not
explored in this study. We quantify the bias  introduced (and show
it is significant) by the naive Euler estimator in Section
\ref{s_res}, but it should be noted that a deterministic
optimization can be completed in less than an 10 minutes (on the
same computing platform). This fact will hopefully motivate
additional research into reliable general deterministic
time-inhomogeneous likelihood expansion techniques.

\subsection{The probability integral transform}
The probability integral transform \cite{diebold} (PIT) is a
powerful tool which can be used in a variety of time series
contexts \cite{hong}.   The technique requires one to have the ability to
evaluate the transition density associated with the assumed model.
In parametric diffusion modeling applications, previous studies
have shown that various  transition density expansions can be
reliably used in place of the exact transition density
\cite{hong,lllv} to approximate the PIT.  The PIT
is a random variable ($Z_m$) which is constructed from an observation pair
and the assumed (or empirically measured) transition density.  The
scalar case is easiest to demonstrate (the multivariate is a
fairly straightforward extension \cite{hong}) and the
transformation is shown below:
\begin{align}
\nonumber  &Z_m:  =\int\limits_{-\infty}^{x_m} p(x_m'|x_{m-1};\theta)dx_m'  \\
\nonumber  &Z_m  \sim q(Z_m;\theta)  \equiv \frac{d \mathbb{Q}(Z_m;\theta)}{dZ_m}  \\
  &x_m  \sim f(x_m|x_{m-1}) \equiv \frac{d \mathbb{F}(x_m|x_{m-1})}{dx_m}
\end{align}

Under a correctly specified model, the $Z_m$'s  are  uniformly
distributed on the interval $[0,1]$ and independent of one another
\cite{diebold}.  The transition density of the true process,
$f(x_m|x_{m-1})$,  may not be adequately described by the assumed
model class (whose transition density here is denoted by
$p(x_m|x_{m-1};\theta)$ ). If this is the case, $q(Z_m;\theta)$
will either not be a uniform density due to the discrepancy
between $f(x_m|x_{m-1})$ and $p(x_m|x_{m-1};\theta)$ and/or the
$\{Z_m\}$ series constructed from the data and the assumed model
will  exhibit serial correlation \cite{diebold}.  A suite of
statistical tests have been developed which exploit these two
facts; here we utilize the $M-$test statistic introduced in
\cite{hong}. To construct it one needs to first construct
 the series $\{Z_m\}_{m=1}^M$ from the series  $\{x_m\}_{m=0}^M$ (for our applications this can be readily done once
 the global spline function is in hand).  The test-statistic ($\bar{M}(m,l)$) is given by the following
 formula:

\begin{equation}
\bar{M}(m,l)\equiv \frac{\Big( \sum\limits_{j=1}^{M-1}
w^2(\frac{j}{p})(M-j)\hat{\rho}_{ml}^2(j) -
\sum\limits_{j=1}^{M-1}
w^2(\frac{j}{p})\Big)}{\Big(2\sum\limits_{j=1}^{M-2}
w^4(\frac{j}{p})\Big)^{\frac{1}{2}}}.
\end{equation}

Where in the above $\hat{\rho}_{ml}^2(j)$ is the sample cross-correlation
 between ${Z}_\tau^m$ and ${Z}_{\tau-j}^l$ (here the superscripts correspond to powers and not to time series indices) and $w(\cdot)$ is a weighting
 function for the lag orders ${j}$ \cite{hong}.
 We make use of  the Bartlett kernel for the weighting function  ($w(z):=(1-|z|)\mathbf{1}_{|z| \le 1}$,
 where $\mathbf{1}_{\mathrm{A}}$ represents the indicator function for
 ``event $\mathrm{A}"$ and $p$ is a prespecified lag truncation parameter).

\section{\label{s_pmf} Established Techniques for Estimating the PMF}

Under appropriate assumptions \cite{jarz97,crooks99}, the
Jarzynski  work relationship (which is one of several recent
fluctuation theorems \cite{crooks99}) allows one to calculate
equilibrium properties from nonequilibrium simulations. It has
received  a lot of attention in biophysics due to its potential
relevance to a wide variety of systems including protein folding,
ion-channels, and wet lab experiments involving optical-tweezers
and atomic force microscopes \cite{hummer_science,hummerPNAS01}.

The purpose of this article is to attempt to develop a technique
which can provide the  large amount of synthetic  data (obtained
through estimating  models based on the genuine process) needed
for reliably estimating an exponential average of a random
quantity.  We will only give the equations needed to construct a
PMF through  exponential averages of the work done on the system.
A detailed treatment of the subject can be found in a variety of
sources including
\cite{jarz97,crooks99,hummerPNAS01,schultenJCP04}.  The Jarzynski
work relation is as follows:
\begin{equation}
\label{e_expINEQ} \Big<\exp{(-\beta W)} \Big>= \exp{(-\beta \Delta
G)}
\end{equation}

Where in the above $W$ is the work done on the system and $\Delta
G$ is the free energy difference between two equilibrium states.
The work introduced changes the system from known equilibrium
conditions characterized by the reaction coordinate(s) in ``state
1" to value(s) associated with ``state 2" (for our problem the
state is simply the numerical values of the linear end-to-end
distance of the deca-alanine molecule at a given time). The
brackets in Equation \ref{e_expINEQ} indicate an ensemble average
over paths \cite{schultenJCP04}. The work done on the system is
obtained by integrating in time; we use the work definition
\footnote{It is common practice in molecular simulation to use a
\emph{single} batch of trajectories to evaluate the PMF at
\emph{several} different time points; we adhere to this practice
despite some of the mathematical technical problems associated
with reusing the paths to evaluate the work integral at different
time points. } given in \cite{schultenJCP04}, that is:
\begin{equation}
W(t):=-v_{\mathrm{pull}}k_{\mathrm{pull}}\int\limits_0^t
\Big(z_{t^{\prime}}-z^{\mathrm{targ}}(t^{\prime})\Big) dt^{\prime}
\end{equation}

There are a variety of techniques for overcoming the difficulty
associated with taking averages of an exponential process.  The
stiff-spring approximation \cite{schultenJCP04} requires  the work
distribution along the SMD path to be approximately Gaussian; this
has the drawback of needing to experiment with different values of
$k_{\mathrm{pull}}$ and $v_{\mathrm{pull}}$ in order to determine
which values this approximately holds for in the particular system
under study (this can be computationally very demanding in some
SMD simulations). The cumulant expansion \cite{schultenJCP04} is
another alternative, but one still needs to make reliable
estimates of the low order moments of an ensemble.  Attempts at
using ML estimation on the ensemble of paths to obtain efficient
estimation has also been explored in previous works
\cite{hooker_MLE}.  It should be stressed that all of the
aforementioned works focused on making best use of a finite set of
work paths.  The surrogate work paths created by our local
diffusion models can be treated by these ensemble methods, but the
focus of our approach is in approximating the statistical
structure of the ensemble of SMD processes so that we can then use
this structure to \emph{create} additional bootstrapped samples
after making observations on a relatively small number of paths.
The eventual \emph{hope} is that the SMD time series  associated
with interesting ``rare events" \cite{jarzysnki06} will correspond
to a small number of coherent structures in our local parameter
spaces and the information in our collection of surrogate models
can be used to approximate the relative frequency and the
distribution of magnitudes that these paths will contribute to the
exponential average (though this is highly optimistic, the results
of this study indicate this is not unreasonable).



\section{\label{s_simdetails} Computational Details}

\subsection{SMD Simulation Details}


Throughout this study the NAMD  program \cite{namdprog}
(\url{http://www.ks.uiuc.edu/Research/namd/}) was used to generate
data.   The pulling speed ($v_{\mathrm{pull}}$) used for all
simulations was $0.01 \mathrm{\frac{\AA}{ps}}$; a step size of 2
fs was used to integrate the equations of motion for $1\times
10^6$ SMD time steps, and the spring constant used for pulling
($k_{\mathrm{pull}}$) was assigned  a value of $7.2$
kcal/mol/$\mathrm{\AA}^2$ in all simulations. Additional
computational details (including parameter files, many of the raw
data sets, etc.) can be downloaded from
\url{http://www.ks.uiuc.edu/Training/Tutorials/science/10Ala-tutorial}
\footnote{The NAMD cv-SMD tutorial (and others related tutorials)
are available at this address .  Coincidentally, we used the time
series (available in MATLAB format at this url) for calibrating
the runs that used a Langevin damping parameter ($\gamma$) of $5
\mathrm{ps}^{-1}$.}.

\subsection{Data Preparation}
The local models were calibrated using the nonstationary time
series that came from the constant velocity SMD simulations. The
net time series generated by the SMD simulations are denoted by
$\mathbf{x} \equiv \{x_i\}_{i=0}^M$.  To use local parametric
models, the observations were screened by inspecting which
observations fell within a window centered at a specified $X_o$
(which is the point at which a local model is desired).  If
$M^{\mathrm{win}}$ (which is random) observations  fell within the
window, then a new
 series $\mathbf{x}^{\mathrm{win}}$ was constructed consisting of the $M^{\mathrm{win}}$ pairs
 $(x_m,x_{m+1})$ where the first coordinate is the observation
 that fell within the selected window.  The screened time series
 was then passed to the scheme that was meant to approximate the log
 likelihood function (in this paper the SML approximation).

 It should be stressed that the underlying data source  only needed
 to be generated \emph{once}.  After the SMD data was created, we were free to try and calibrate any number of
 local models around the time series (this task can easily
 be distributed across independent processors).  After using the observations to estimate a
   global model, the
 same SMD observations can be used again to determine quantitatively how
 well the data matches the calibrated stochastic model.

\subsection{Estimation Details}
The parameters were obtained by running $50$ Nelder-Mead \cite{nm}
searches where the Brownian paths needed for the SML routine were
generated once per search.  This was done to minimize the
variance associated with using a noisy likelihood expansion
technique (each parameter search used a different (random) initial
parameter guess).  After determining the optimal parameter vector,
the parameters $T_{\mathrm{sml}}$ and $N_{\mathrm{sml}}$ were
increased to determine if the optimal parameter vector changed
significantly.  For the results shown, the values
$T_{\mathrm{sml}}=50$ and $N_{\mathrm{sml}}=350$  gave results
that did not change significantly after increasing the SML
likelihood expansion parameters.

\section{\label{s_res} Results }

\subsection{Estimation Results}

\begin{figure}[h]
\includegraphics[angle=0,scale=.40]{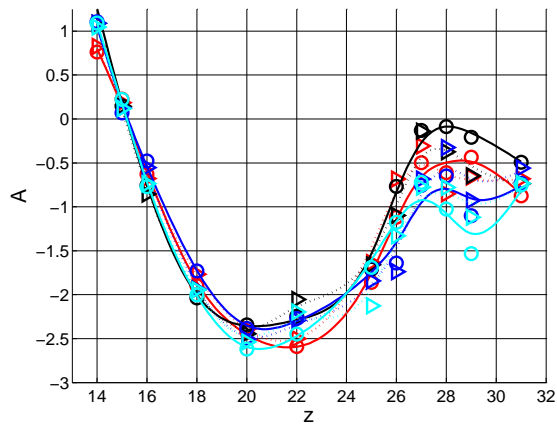}
\caption{\label{fig:PARdriftDS1A} The global drift function
estimated by a smoothing spline fit of the constants measured from
sampling of the time series coming out of the constant velocity
SMD simulation. The dashed lines correspond to the spline obtained
(constructed using the estimated local parameter values indicated by
symbols) by sampling the data every 50 SMD times steps ($0.1
\mathrm{ps}$) and the solid lines correspond to the parameters
estimated using the same data sampling every 150 SMD times steps
($0.3 \mathrm{ps}$). All simulations used a damping coefficient
$\gamma=5$ $\mathrm{ps}^{-1}$ for the Langevin
heat bath. } 
\end{figure}

\begin{figure}[h]
\includegraphics[angle=0,scale=.40]{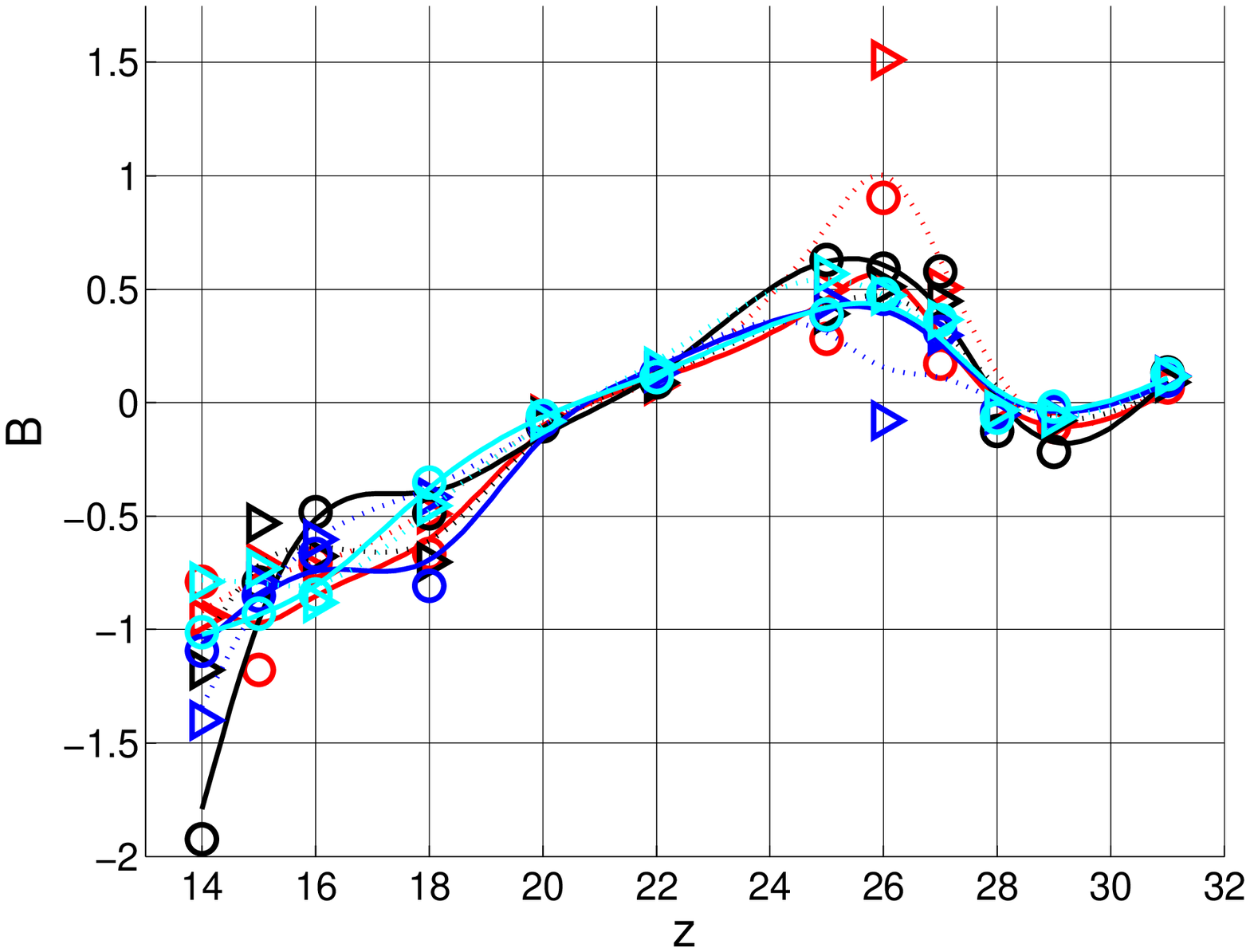}
\caption{\label{fig:PARdriftDS1B} The ``linear sensitivity" (as
measured by piecing together the $B$ parameters through  smoothing
splines) of the global drift function. For computational details,
see caption in Figure \ref{fig:PARdriftDS1A}.}

\end{figure}

\begin{figure}[h]
\includegraphics[angle=0,scale=.40]{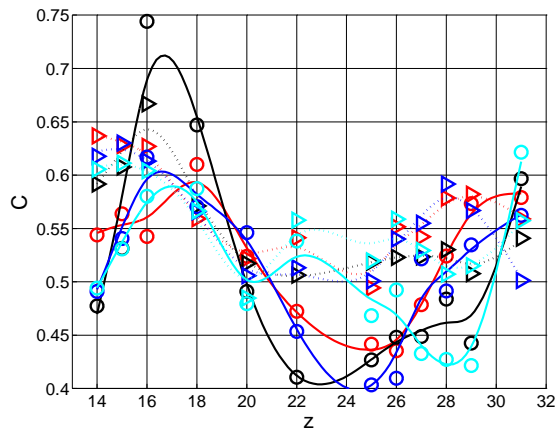}
\caption{\label{fig:PARdriftDS1C} The global diffusion function
estimated by a smoothing spline fit of the constants. For computational details see
 caption in Figure \ref{fig:PARdriftDS1A}.} 
\end{figure}

\begin{figure}[h]
\includegraphics[angle=0,scale=.40]{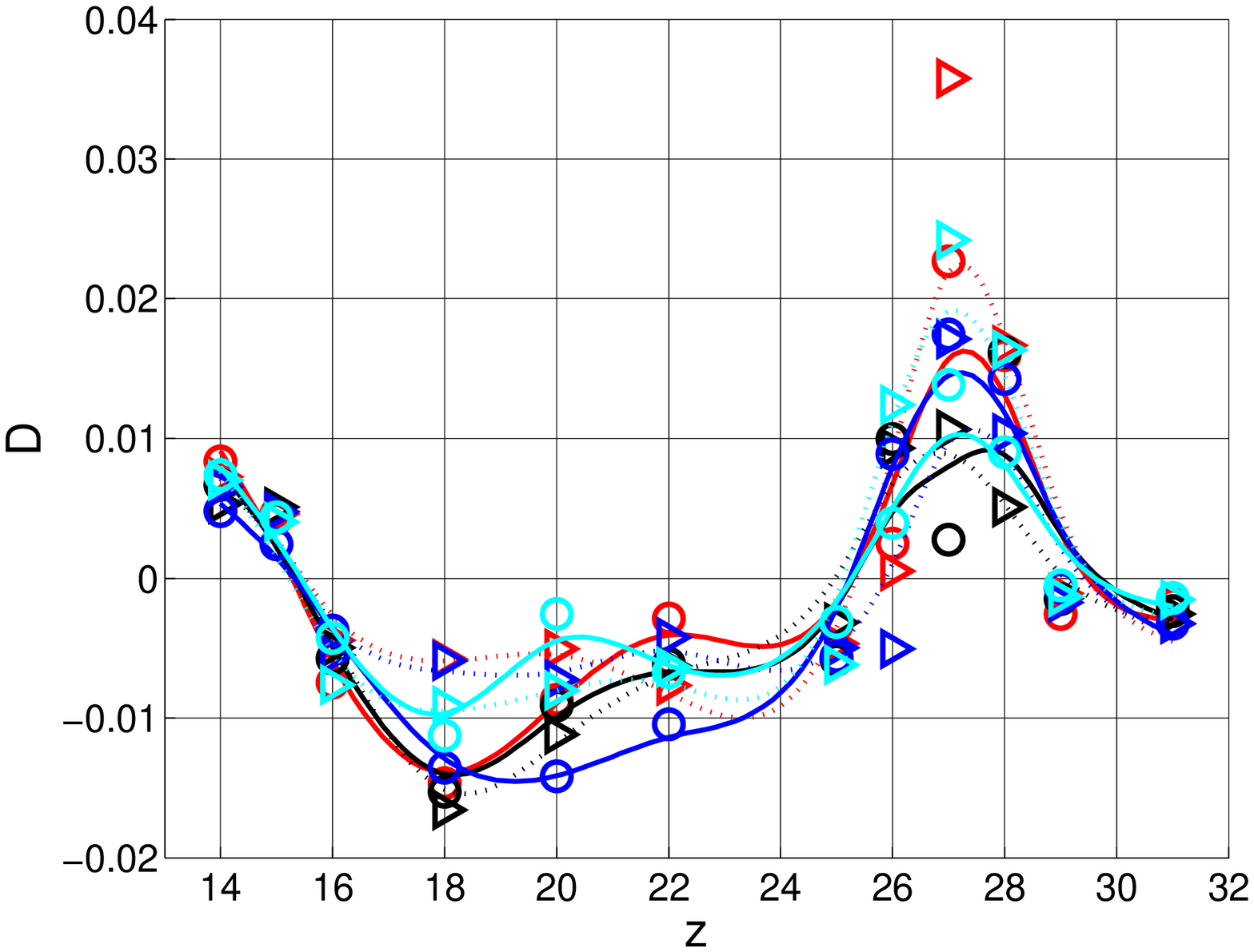}
\caption{\label{fig:PARdriftDS1D} The ``linear sensitivity" (as
measured by piecing together the $D$ parameters through smoothing
splines) of the global diffusion function.  For computational
details, see caption in Figure \ref{fig:PARdriftDS1A}.}
\end{figure}

Figures \ref{fig:PARdriftDS1A}-\ref{fig:PARdriftDS1D} plot the
various local model parameters associated with the assumed local
model estimated at $12$ different points using a window width of
$2 \mathrm{\AA}$ centered at the values of $z$ indicated by the
presence of a symbol.  The lines connecting the model parameters
were constructed using MATLAB's smoothing spline function
\url{csaps}.  In all of the aforementioned plots, the dashed lines
correspond to parameters estimated from data that was sampled from
the NAMD simulation every 50 time steps (this resulted  in
$\approx 2100$ observations per window on average) and the solid
lines correspond to the parameters estimated from data that was
sampled from the NAMD simulation every 150 time steps (this
resulted  in $\approx 700$ observations per window on average).
The fact that the Langevin thermostat was used to regulate
temperature in these simulations should also be explicitly noted
(both cases plotted in Figures
\ref{fig:PARdriftDS1A}-\ref{fig:PARdriftDS1D} used  $\gamma=  5
\mathrm ps^{-1}$). In order to construct the global diffusion
model from this data, we used the splines that came from piecing
together the drift and diffusion constants ($A$ and $C$). This was
done because these parameter estimates were observed to have less
relative noise associated with them due to estimation error in
most cases (e.g. see Table \ref{tab:pardist}).  The ``linear
sensitivity" ($B$ and $D$) coefficients are still useful for
determining features of the global nonlinear functions.  For
example note how the location of the local extrema in the global
spline function can be predicted from the zero crossings of the
estimated  linear sensitivity components.  One observes that the
qualitative features of the drift and diffusion coefficient
functions are similar, but the parameters estimated for each
individual SMD realization appears to come from a slightly
different function (finite sample estimation noise alone should
not connect smoothly if the true process came from  a single
scalar diffusion) . Each function appears to be a smooth
distortion of one underlying (unknown) base function.   It should
also be noted that the noise parameter is significantly affected
by the sampling frequency whereas the drift function appears to be
relatively insensitive to the sampling frequency. Possible causes
for the discrepancy  will be discussed in the goodness-of-fit
results reported later.  For now it should be noted that the
scalar reaction coordinate description of the system is itself a
drastic oversimplification of the system; in addition it is also
well known that the dynamics of a reaction coordinate that
accounts for only positions and not velocities is not necessarily
Markovian \cite{zwanzig}.   A SMD simulation that is carried out
in the presence of explicit solvent has noise which is typically
significantly correlated for the observation frequency we use to
estimate the parameters of a diffusion  model with (though the
artificial ``solvent" in our data was modeled using a Langevin
thermostat which helped in reducing this type of correlation).
 If the true dynamics of the reaction coordinate are
 not Markovian then one should be concerned about the errors introduced
 when one tries to use the Jarzynski work relationship to construct a PMF
 using a Markovian surrogate process.  The tests given in
\cite{hong} quantitatively test the statistical validity of the
simplifying assumptions we impose on the proposed stochastic model
and even offer insight into the inadequacies of our simplified
description and Section \ref{s_extension} discusses one possible
way in which we can introduce a more realistic noise process.

\begin{table*} [h]
\caption{\label{tab:pardist}MC Parameter Distributions - Local
model parameters estimated from 100 realizations of a
\emph{single, known} SDE diffusion whose coefficient functions
were determined by  estimation using data from a \emph{single} SMD
simulation (using data sampled every 50 SMD steps with $\gamma=5
\mathrm ps^{-1}$).  The window size was 2$\mathrm \AA$ centered at
two different values of  $\mathbf{X_o}$ (the row with the $\mathbf{X_o}$ reported
contains the \emph{known} ``local values" of the spline
functions).  }
\begin{ruledtabular}
\begin{tabular}{lcccc}

& $A$ & $B$ & $C$ & $D$  \\ \hline \hline

$\mathbf{X_o=14}$ & $8.409\times10^{-1}$  &    $-7.144\times10^{-1}$ & $6.376\times10^{-1}$  &  $-6.053\times10^{-3} $  \\ \hline
Mean Euler & $9.024\times10^{-1}$ &  $-8.281\times10^{-1}$ &   $5.178\times10^{-1}$ & $-5.803\times10^{-3}$  \\ \hline
 Std. Euler & $9.965\times10^{-2}$ &  $1.648\times10^{-1}$ &   $7.158\times10^{-3}$ & $1.830\times10^{-3}$  \\ \hline
Mean SML   & $8.161\times10^{-1}$ &  $-6.827\times10^{-1}$ &   $6.627\times10^{-1}$ & $-5.981\times10^{-3}$  \\ \hline
Std. SML   & $1.209\times10^{-1}$ &  $1.767\times10^{-1}$ &   $1.857\times10^{-2}$ & $1.842\times10^{-3}$  \\ \hline \\ \hline
$\mathbf{X_o=22}$ & $-.2501\times10^{1}$  &    $9.051\times10^{-2}$ & $5.322\times10^{-1}$  &  $-4.281\times10^{-3} $  \\ \hline
Mean Euler & $-.2474\times10^{1}$ &  $9.685\times10^{-2}$ &   $4.610\times10^{-1}$ & $-3.934\times10^{-3}$  \\ \hline
 Std. Euler & $1.369\times10^{-1}$ &  $4.444\times10^{-2}$ &   $7.290\times10^{-3}$ & $2.105\times10^{-3}$  \\ \hline
Mean SML   & $-.2465\times10^{1}$ &  $8.613\times10^{-2}$ &   $5.550\times10^{-1}$ & $-4.165\times10^{-3}$  \\ \hline
Std. SML   & $1.673\times10^{-1}$ &  $2.523\times10^{-2}$ &   $1.514\times10^{-2}$ & $1.360\times10^{-3}$  \\ \hline

\end{tabular}
\end{ruledtabular}
\end{table*}

Table \ref{tab:pardist} reports Monte Carlo parameters estimates
obtained from  local estimation using synthetic data (from a
single global SDE) that was created  using a trajectory of the 
$\gamma =5 \mathrm ps^{-1}$  NAMD runs (sampling every 50 SMD
steps) and then using the estimated \emph{single} (genuine) SDE to
generate 100 paths with the corresponding (\emph{exactly known})
global spline coefficient functions. We see that for the sampling
frequency used that the estimators produce significantly different
results. Deterministic likelihood approximations are attractive
because they can typically be executed with much less CPU time,
but obtaining a general (reliable) deterministic expansion method
for a wide class of multivariate, time-inhomogeneous,
state-dependent noise cases appears to be beyond our current
analytical capabilities (however the SML easily can accommodate
this case at the cost of being computationally inefficient).

\subsection{Synthetic Work and Approximating the PMF}

\begin{figure}[h]
\includegraphics[angle=0,scale=.40]{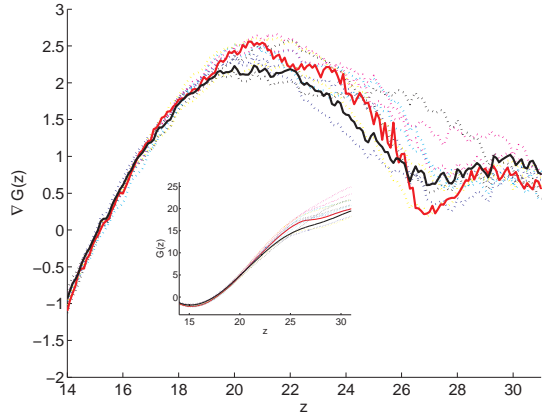}
\caption{\label{fig:PMFest} The estimated gradient of the PMF obtained by running 100 (bootstrapped) copies of the spline estimated from one SMD trajectory (dashed lines) and that obtained by pooling the work trajectories (solid black) and the PMF obtained through a benchmark reversible pulling (solid red) experiment \cite{schultenPMF} (the inset displays the corresponding PMF).}
\end{figure}

The SDEs corresponding to the  splines plotted in the previous
figures were then used to generate ``synthetic" work paths.  Each
of the $10$ estimated SDEs were used to create 100 synthetic
realizations of the work paths associated with the SMD stretching
experiment. Figure \ref{fig:PMFest} plots the gradient of the
estimated PMF using the various work paths and directly using the
relation given in Equation \ref{e_expINEQ} (the inset plots the
raw PMF).  The dashed lines in the plots correspond to the PMF
estimated from 100 copies estimated along a single (estimated)
diffusion model; the solid black line corresponds to that of
pooling the work paths from all synthetic trajectories together
and then estimating the resulting PMF through evaluating the
empirical exponential average and the solid red line plots the
``exact" results \cite{schultenPMF}.

 We could easily
increase the number of synthetic paths in an attempt to (at a
relatively cheap computational cost) to create a larger number of
synthetic trajectories in an attempt to get the large sample sizes
needed to safely calculate an exponential average; however, a
major point of this paper is that it more important to
characterize the distribution of \emph{SDE coefficient functions}
than it is to sample the work paths that come from a small set of
estimated SDEs. How this observation can possibly be exploited is
discussed in Section \ref{s_extension}.

\begin{figure}[h]
\includegraphics[angle=0,scale=.40]{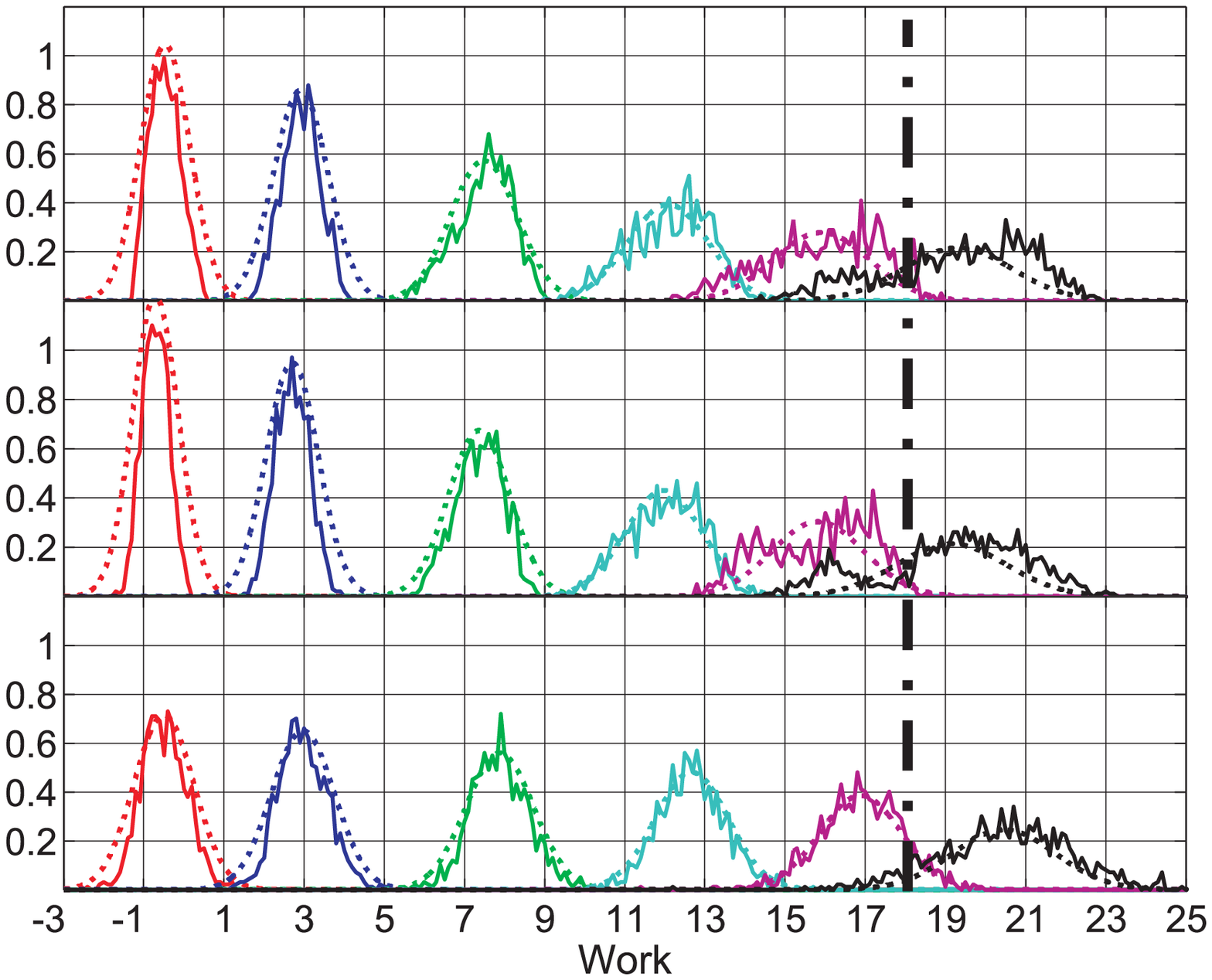}
\caption{\label{fig:workdist} The empirically measured work
probability densities (the Gaussian distribution corresponding to
the empirically measured mean and standard deviation is also
plotted using dotted lines in all figures above).  The top plot
shows the results corresponding to running 100 (bootstrapped)
copies for each of the 10 estimated global surrogate diffusion
models obtained by sampling the 10 SMD trajectories every 150 SMD
steps ($0.3 \mathrm ps$) ; the middle those corresponding to
sampling every 50 SMD steps ($0.1 \mathrm ps$); the bottom plot
are those obtained by running 1000 genuine SMD trajectories.  All
simulations used a damping coefficient $\gamma=5$
$\mathrm{ps}^{-1}$.  The work distributions plotted correspond to
the distributions obtained at times ($.4, .6, .8, 1.0, 1.2, 1.5$)
ns in all cases.}
\end{figure}

The top two plots in Figure \ref{fig:workdist} show the work
distributions associated with the $1000=10^{locmod}\times
100^{copies}$ synthetic paths and the bottom plot shows the work
distribution corresponding to the same number of genuine SMD
simulations. For each distribution plotted, the corresponding
normal distribution (using the empirically measured mean and
standard deviation) is plotted as well (using dashed lines).  This
was done because the stiff-spring approximation
\cite{schultenJCP04,ioan} requires a normal distribution of work.
A relatively small deviation from the normality assumption using
an exponential average to calculate the free energy can greatly
affect the PMF calculation
 (recall our methodology does not require a
normal distribution of work).

The thick dashed-dotted line in Figure \ref{fig:workdist}
represents the ``reversible work" \cite{crooks99} corresponding to
the SMD process at $1.5 \mathrm{ns}$ (this time is associated with
a target end-to-end distance $z=28 \mathrm{\AA}$). The reversible
work was determined from the ``exact" PMF \cite{schultenJCP04}.
Any value of work (in the histograms corresponding to $1.5
\mathrm{ns}$) to the left of this line is considered to be an
important rare event (interpreted physically, this corresponds to
a violation of the second law of thermodynamics at the atomistic
scale \cite{crooks99}). Note that the differences in the shapes of
the tails of the actual work distribution and those of the
synthetic work distributions are significant; this accounts for
the discrepancy in the PMF estimated using synthetic work paths at
this state value. It should also be mentioned that the work
distribution at this point does include an \emph{accumulation} of
errors from previously visited state points. Observe how the
shapes of the synthetic distributions deviate more from those of
the genuine SMD simulations as time progresses (errors are
introduced both by finite sample estimation errors and
inadequacies of the diffusion model). The biggest source of error
in the estimation of the PMF is likely due to the fact that all
bootstrapped samples were given equal weight (bootstrapping a
surrogate model obtained using data corresponding to a rare event
explains the appearance of multiple modes in the $1.5 \mathrm{ns}$
synthetic work histograms). Possible remedies to these and other
problems are discussed and demonstrated on a toy model in Section
\ref{s_extension}.


\subsection{Testing the validity of the diffusion approximation}

In the previous section we saw that the PMF estimated from a small
set of synthetic SDE trajectories could faithfully reproduce the
PMF through approximating the work paths  needed for the
exponential average used in the Jarzynski nonequilibrium work
relationship (in the relatively ``simple" deca-alanine system
studied).  This may just be due to the fortunate fact that in the
system studied that the work relationship has a certain
``robustness" to the process that generates the work paths. Here
``robustness" refers to a situation where a large collection of
significantly different stochastic processes generate  similar
work distributions and the PMF estimated is nearly independent of
the differences in the work distributions that come from the
various processes \footnote{Inspection of Equation \ref{e_expINEQ}
shows that this is possible for a wide variety of work
distributions and the ``similarity" in the work distribution may
not be too stringent outside of the region containing interesting
rare events.}.

\begin{figure}[h]
\includegraphics[angle=0,scale=.40]{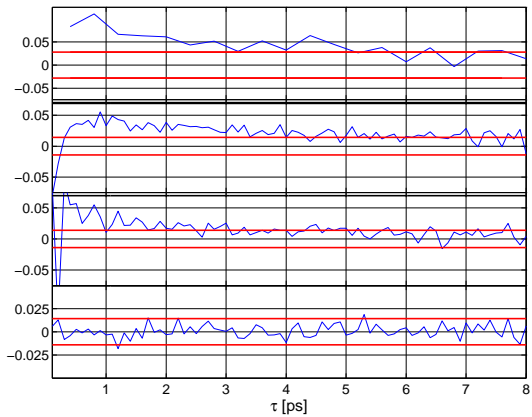}
\caption{\label{fig:AC} The sample correlation
 function between $Z_{m}$ and $Z_{m-j}$ (Note: $Z$ corresponds to the PIT rv,
 not the end-to-end distance $z$).
  The top figure corresponds to the PIT sequence obtained by using a
  genuine SMD trajectory sampled
  every 150 SMD time steps using $\gamma=5$ $\mathrm{ps}^{-1}$ and the transition
  density corresponding to the  global model obtained using the
  same trajectory ;
  the second plot from the top to that of using data from a genuine
  SMD
  trajectory sampled every 50 SMD time steps using $\gamma=50$ $\mathrm{ps}^{-1}$ and a
   transition density corresponding to a surrogate
  model using the same  trajectory
  ;
  the second plot from bottom to that of using a genuine SMD trajectory
  sampled
  every 50 SMD time steps using $\gamma=5$ $\mathrm{ps}^{-1}$ and a transition density
  obtained from the same  trajectory ; and the bottom
   to that of using a synthetic SDE trajectory (different from the trajectory used for estimation)  with a SML approximation of the transition density that actually generated the data (obtained  using
   model parameters from estimating along a SMD trajectory sampled
   every 50 SMD time steps using $\gamma=5$ $\mathrm{ps}^{-1}$).}
\end{figure}

\begin{figure}[h]
\includegraphics[angle=0,scale=.40]{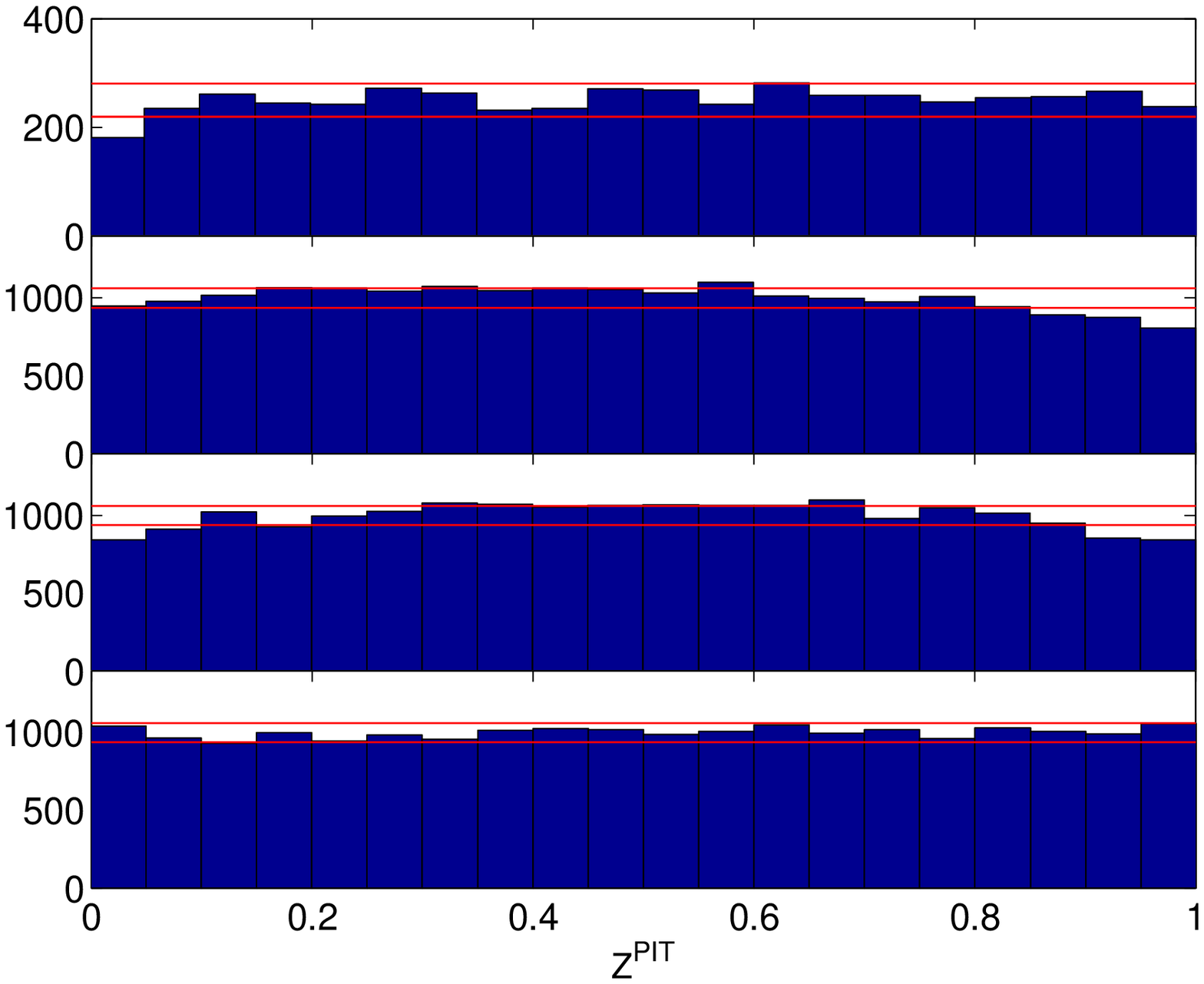}
\caption{\label{fig:Zhist} The histogram of $Z_{m}$ corresponding
to the global surrogate model . The top plot corresponds to the
first SMD trajectory sampled every 150 SMD time steps using
$\gamma=5$ $\mathrm{ps}^{-1}$; the next to that the case of
sampling every 50 SMD time steps using $\gamma=50$
$\mathrm{ps}^{-1}$;the third plot to  the case of sampling every
50 SMD time steps using $\gamma=5$ $\mathrm{ps}^{-1}$; and the
bottom  to a synthetic SDE trajectory (with splines constructed
using the first SMD trajectory sampled every 50 SMD time steps and
$\gamma=5$ $\mathrm{ps}^{-1}$).  }
\end{figure}

\begin{figure}[h]
\includegraphics[angle=0,scale=.40]{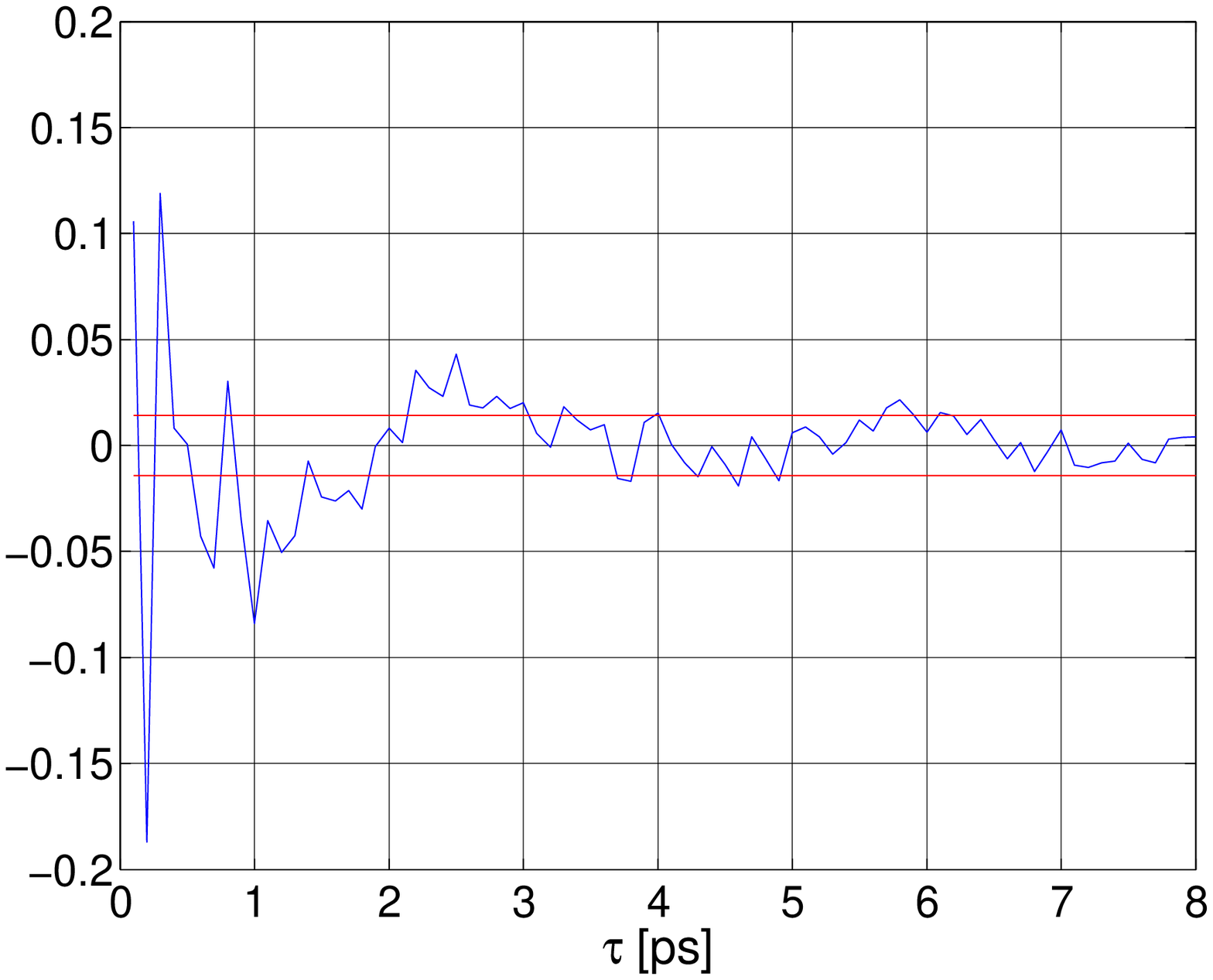}
\caption{\label{fig:udAC}   The autocorrelation of the PIT
associated with actual SMD data corresponding to a transition
density obtained by estimating the parameters sampling every 50
SMD time steps and using $\gamma=0.1$ $\mathrm{ps}^{-1}$ (using the same trajectory). See
Figure \ref{fig:AC} for additional details.  Note the more
oscillatory nature of this empirical autocorrelation. }
\end{figure}

Here we carry out goodness-of-fit tests that investigate
quantitatively how well the diffusion approximation is given the
global splines estimated from a sequence of local models and the
actual SMD data.  The results depend heavily on the sample size,
sampling frequency and on the statistical validity of the
diffusion approximation. Figures  \ref{fig:AC} and \ref{fig:Zhist}
plot the sample correlation and histogram of $\{Z_{m}\}$
associated with one realization of the SMD process (using the
estimated global nonlinear diffusion model with the SML expansion
of the transition density).  The red lines in both figures
correspond to $\pm 2\sigma$ of the asymptotic distributions
associated with a correctly specified model.  Four different cases
are plotted: one using the data and splines estimated from
sampling from the SMD process every 150 steps with $\gamma=5$
$\mathrm{ps}^{-1}$, another sampling every 50 steps with
$\gamma=50$ $\mathrm{ps}^{-1}$, the third sampling every 50 steps
with $\gamma=5$ $\mathrm{ps}^{-1}$ and the fourth that associated
with using a genuine SDE with the splines estimated from the first
trajectory sampled every 50 steps with $\gamma=5$
$\mathrm{ps}^{-1}$. The overall shape of the histogram for the
empirical SMD cases does not appear too bad given the size of the
time series used (recall all SMD simulations contained
$1\times10^6$ time steps).  The sample autocorrelation function
indicates that there is significant serial correlation in the
$\{Z_m\}$ series when using the actual  SMD data.
The results from the synthetic SDE trajectory are shown because
the SML expansion may cause measurable errors in the PIT (the
results indicate that these approximation errors are negligible in
comparison to the imperfections of the assumed model). It should
be noted that although we only show results from a single
trajectory, the results for the other trajectories are
qualitatively similar for the paths observed (this has interesting
implications, see Section \ref{s_extension}) . It is also
interesting to observe the results obtained when one estimates the
parameters using a lower damping coefficient and then analyzes the
associated PIT series. In Figure \ref{fig:udAC} we estimate a
model by sampling a SMD trajectory every 50 steps and use
$\gamma=0.1$ $\mathrm{ps}^{-1}$ (it is likely that the system is
underdamped using this damping coefficient and sampling
frequency); note that the autocorrelation function clearly
displays a more oscillatory response in comparison to  the higher
values of $\gamma$.

\begin{table*} [h]
\caption{\label{tab:Mvalues}Computed M-test Statistic (``ds" denotes the down sampling rate;
 e.g. ds=50 indicates that the time series was constructed using the output of the SMD simulation
 recorded every 50 steps) and the ordered pair in the column indicates $\bar{M}(m,j)$.}
\begin{ruledtabular}
\begin{tabular}{lcccccc}

Case & $\mathbf{(1,1)}$ & $\mathbf{(2,1)}$ & $\mathbf{(1,2)}$ & $\mathbf{(2,2)}$ & $\mathbf{(3,3)}$ & $\mathbf{(4,4)}$  \\ \hline \hline
Empirical ($\gamma=5 \mathrm{ps}^{-1}$, ds=150) & 20.5337 &  19.3370 &  19.9771  & 18.9426 &  16.6009 &  14.4330  \\ \hline
Empirical ($\gamma=50 \mathrm{ps}^{-1}$, ds=50) & 84.0644 &  80.1934 &   80.8452  &  81.6811 &  78.1994 &   74.2394  \\ \hline
Empirical ($\gamma=5 \mathrm{ps}^{-1}$, ds=50) & 92.2045  & 85.0764 &  86.0400 &  81.7107  & 73.3389  & 66.4957  \\ \hline
Synthetic ($\gamma=5 \mathrm{ps}^{-1}$, ds=50) & -0.3584 &  -0.1151 &  -0.6810 &   -0.6483 &  -0.9822 &    -1.3087 \\ \hline
Uniform i.i.d. rv   & -0.7169 &  -0.8644 &  -0.8615 &  -0.8979 &  -0.9049 &  -0.8484 
\end{tabular}
\end{ruledtabular}
\end{table*}

Table \ref{tab:Mvalues} offers further insight into the model
inadequacies. The first four columns report the computed
$\bar{M}(j,m)$ statistic for a single realization for the various
cases listed in the table . The final column plots the
test-statistic obtained when a truly uniform (on $[0,1]$)
independent and identically distributed (i.i.d.) sequence of
random variables is used (using the sample size corresponding to
sampling every 50 SMD steps) to construct $\bar{M}(j,m)$.  If the
model is correctly specified, this statistic should be normally
distributed with unit variance. We can obviously strongly reject
all of the ``empirical" models (this rejection should not be too
discouraging given the fact that we have a fairly large time
series of observations which facilitates rejecting the null
 hypothesis).

  These results suggest that some of the
fast time scale motions of the underlying N-body process are still
statistically measurable at the sampling frequency used. The fact
that the $\gamma=50 \mathrm{ps}^{-1}$ case only appears to
 be marginally better than the corresponding  $\gamma=5 \mathrm{ps}^{-1}$ case   indicates that our
overdamped approximation for the observation frequency used is
probably not the major problem in our model (this is further
supported by the empirical evidence in Figure \ref{fig:udAC}). The
 $Z_m$ series (for the cases where $\gamma \ge 5 \mathrm{ps}^{-1}$) appears to be negatively correlated for a
very short time window and for longer times it has a positive
correlation that decays slowly. These phenomena are more than
likely caused by the noise introduced by the fast vibrational
motion and the slower long range Lennard-Jones type interactions
between nonadjacent atoms of the deca-alanine molecule
(respectively). The fact that $\bar{M}(1,1)$ is the greatest in
all of the ``empirical cases" strongly suggests that something is
askew with the drift; it is likely that the dynamics of a scalar
reaction coordinate are either non-Markovian and/or the drift
coefficient is more oscillatory than we assumed (our local models
only yield smooth local functions).  In some cases it may be
possible to reduce the error associated with the former
possibility by adding another reaction coordinate into the model
and/or sampling the SMD data less frequently; the latter
possibility can be tested by using smaller windows for the local
model. Unfortunately if a very small
 window is used, the behavior of the small
 sample parameter estimates may become
 sporadic (increasing the sampling frequency
  is also problematic due to the fact that
  the Markovian assumption on the dynamics of the reaction coordinate typically breaks
  down at some limit of sampling frequency \cite{elansary,zwanzig,stuart06}).










\section{\label{s_extension} Discussion and Possible Extensions}
We have already mentioned  that in certain systems, the Jarzynski
work relationship may not be computationally practical when  work
paths associated with realizations that have a low probability
 cause a large influence on the PMF computed from a
 finite sample of realizations \cite{jarzysnki06}.
 Under these circumstances, the number of realizations
 needed to get a good estimate of the exponential
 average used in the work relationship is too
 large in many systems of interest. This fact fuels interest
 in methods which can infer the basic statistical
 structure from a small number of actual SMD realizations.

 In this paper, we  demonstrated a method by which one
 can approximate the dynamics of a  reaction coordinate associated with a
  \emph{single} realization
 by fitting a  diffusion SDE (this is then repeated using different SMD realizations).
  The advantage this offers is that
 one can associate a diffusion model with  a
  configuration  drawn from a microscopic ensemble assumed to be at
  equilibrium (recall the Jarzynski work relationship requires the initial configuration to be drawn
 from  equilibrium distribution \cite{crooks99}).  The diffusion model estimated (which will also depend on
 the SMD parameters) can then be ``bootstrapped" by
  running the SDE using different driving Brownian paths in an
   effort to increase work path sampling.  The basic idea being that an interesting
   rare event may be associated with the equilibrium configuration
   used
   (or a minor perturbation of it),
    we just did not observe the rare event in the single realization
    drawn.

     The bootstrapped samples can even be given a loose physical interpretation:
       Running multiple surrogate simulations using different
       Brownian paths (using a single, \emph{fixed}, estimated  diffusion)
        corresponds roughly  to multiple realizations from a stochastic
        process that describes the dynamics associated  with certain features of \emph{one}
        configuration  (e.g. a configuration obtained
        by recording the positions
        of all of the atoms in the deca-alanine molecule from a previous run, but
        assigning new velocities). The ``noise" of the
        process is meant to represent the influence that various items have  on the dynamics of the low-dimensional
         reaction coordinate.  Examples of possible noise sources include the ``internal"
        neglected degrees of freedom of the molecule of interest (e.g. the dihedral
        angles),
        the effects of different initial velocity distributions, and different solvent configurations
        \footnote{In reality, this type of compartmentalization  is
         oversimplified.    Different solvent configurations with all other atomistic
         positions and velocities fixed can, in principle, result in systematically
         different diffusion models; however there are usually \emph{some} configurational
         details that can be adequately accounted for by completely random terms.
         The bootstrapped samples are meant to capture the effects of these details. }.

This paper also claimed that one might have a collection of
diffusion models associated with the ``pulling dynamics" in
certain systems and that the collection has a structure that can
possibly be exploited. One can hope that in general systems the
dynamics
 of a good set of reaction coordinates (whose selection is not always trivial) associated with a
 single realization from the SMD simulation
 may be describable by a diffusion SDE with
 deterministic coefficients, however the
 coefficient functions of the
 SDE may depend very heavily on the details of the
 underlying initial atomistic configuration
 drawn.
 If this is the case, then one may be
 interested in characterizing the
 distribution of coefficient
 functions in an effort to improve
 the PMF estimation by using the
 Jarzynski work relationship with
  synthetic work paths.
  If one somehow knew the distribution of curves,
  then one could appeal to  tools associated with decision theory
  \cite{berger}
  to improve work path sampling
  by using bootstrapped samples  from
  the collection of associated surrogate diffusion models.

To make a concrete demonstration of these ideas, we turn to a toy
model. Suppose the dynamics of a scalar reaction coordinate of
interest are given by:
\begin{equation}
dX_t=\kappa_i(\alpha_i-X_t)dt+\sigma_i \sqrt{X_t}dW_t
\label{eq_cir}
\end{equation}
Where the parameter vector $\Theta_i :=[\alpha_i,\kappa_i,\sigma_i]$ is
distributed according to the normal law $N(\mu,\Sigma)$ where $\mu=[\alpha,\kappa,\sigma]$ is a
 deterministic vector and $\Sigma$ is a deterministic constant symmetric positive definite matrix.
  Suppose we calculate the ``exact" PMF associated with this toy model by sampling 1000 realizations obtained
  by first drawing a $\Theta_i$ (using an
  initial state value from the invariant distribution
  associated with Equation \ref{eq_cir} using $\Theta_i =\mu$) and then carrying out a
   constant velocity pulling experiment;
    we then ask: ``What  can we do to improve the PMF estimation using small sample
    sizes \emph{knowing} a priori that the we have multiple underlying models
    that come from a \emph{known} distribution of model parameters?"

\begin{figure}[h]
\includegraphics[angle=0,scale=.40]{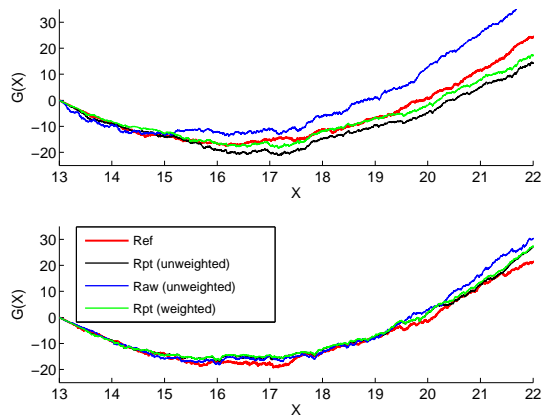}
\caption{\label{fig:CIRtoy}  The top figure was obtained using 10
genuine SMD trajectories and the bottom plot was obtained using
100 trajectories.  The difference between the curves is discussed
in section \ref{s_extension}. }
\end{figure}

In Figure \ref{fig:CIRtoy} the ``exact" and estimated PMFs are
plotted using 10 and 100 genuine SMD (top and bottom plots
respectively) paths. The curve labelled ``Ref" contains the PMF
estimated using 1000 genuine trajectories. The curve labelled
``Raw"  uses only a subset (10 or 100) of the original SMD paths
to compute the PMF using a standard empirical average of the
exponential random variables. The curve labelled ``Rpt
(unweighted)" plots the result obtained by noting the $\Theta_i$
and initial condition of each of the ``Raw" trajectories and
running additional simulations to create more work paths (100 and
10 respectively).  The PMF is then computed by taking standard
exponential averages with the additional trajectories. In the
curves labelled ``Rpt (weighted)" we do the same except in the
exponential average we assign weights determined by the observed
$\Theta_i$ and its known distribution (the density determines the
relative weight and this is used to create a linear weighting
coefficient).  It should be noted in this toy example we assume
perfect parameter estimation (which in practice we do not usually
have the luxury of) and exact knowledge of the underlying
distribution associated with the $\Theta_i$ random variables
(which we will likely never have).  The weighting scheme used is
done for illustrative purposes only (in \cite{hooker_MLE}
``optimal" weighting schemes are discussed)

 The aim of this example was to demonstrate that in small
samples, there can be a significant benefit to estimating the
parameters associated with a single trajectory and rerunning a
synthetic trajectory to aid in (approximate) work path sampling.
The demonstration also showed that a simple  weighting scheme
could be of assistance in a problem that satisfies our assumptions
(in complex systems we can only hope that the simple assumptions
made about the toy model approximately carry over).  The example
also shows that the largest benefit would come from a sampling
technique that would allow us
 to accurately draw from the distribution associated with the
$\Theta_i$ random variables and then run a larger batch of
synthetic SDEs.  Of course estimating this multivariate
distribution with a small sample size with no prior information is
a \emph{very} hard task, but in the deca-alanine system studied
it \emph{appears} that the local parameters are all drawn from a
single connected region in parameter space which may be  well
approximated by a simple distribution like a multivariate normal
(this is just speculation).

In the future, the ideas laid out in this  study should be applied
to other more complicated systems to see if the coefficient
functions of the SDE used to approximate the dynamics of the
reaction coordinate seem to come from a small family of functions.
In complicated systems, we will probably  not be lucky enough to
observe a single family of functions.  The hope is that a small
number of easily discernable families emerge and that these
families can be approximated by employing the help of those
involved with empirical Bayes \cite{berger} and growth curve
analysis \cite{ramsay}.  An interesting twist encountered here is
in how to utilize the information obtained about neighboring local
models into making decisions about a specific local model.

Success in this type of research endeavor  would possibly allow us
to generate the large number of samples needed to accurately
sample rare work events by using bootstrapped samples from a
family of synthetic surrogate stochastic processes. This may
result in an approximation scheme that significantly reduces the
computational load typically needed  to generate the work paths by
using standard SMD simulations (furthermore it is likely that the
local models and parameters characterizing the local models depend
smoothly on system parameters like temperature).

 Characterizing the distribution of curves may be too computationally difficult in some situations.
 However the structure contained in the models we fit can be utilized to test a variety of
 other hypothesis given the model and the data (this information can then be used
 to gain useful quantitative insight about the system). For example the Markov property can be
 tested utilizing the transition density of the assumed diffusion model (or any other
 stochastic process that allows one  a method to evaluate the associated transition
 density).
  Also, the goodness-of-fit improvement obtained  by adding other reaction coordinates or
  a more complicated
  model of the heat bath into the surrogate model can be quantified by using the basic idea
  behind the approach presented (for example in the deca-alanine example we may want to
  monitor the dynamics of the radii of gyration of the molecule in addition to the
  linear end-to-end distance and/or use a jump process to model the breaking of important hydrogen
 bonds). The type of diagnostics discussed above
 can be used to (quantitatively) approximate the timescale
  at which a diffusion model is valid and/or determine the smallest number
  of reaction coordinates needed to
  write a meaningful coarse-grained dynamic model.

\begin{figure}[h]
\includegraphics[angle=0,scale=.40]{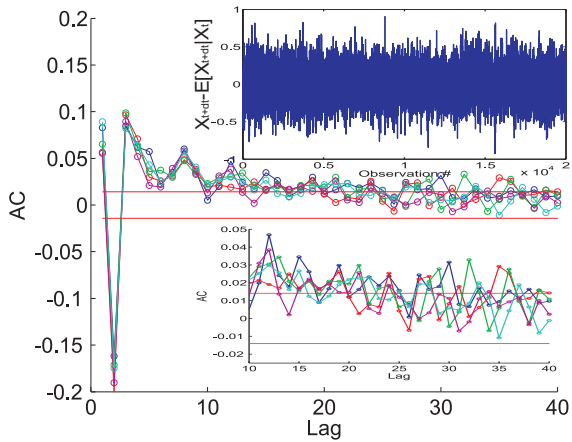}
\caption{\label{fig:ACcombo}  The empirically measured autocorrelation function is plotted for the conditional residuals := $X^i_{t+\Delta t}-\mathbb{E}^{\mathbb{P}_{\theta_i}}[X^i_{t+\Delta t}|X^i_{t}]$ where the superscripts are meant to distinguish between  diffusion models and paths associated with different SMD realizations (here results from five different SMD realizations are plotted).  The straight lines correspond to the $95\%$ confidence bands under the assumption of no correlation.  The top inset plots the time series of $X^i_{t+\Delta t}-\mathbb{E}^{\mathbb{P}_{\theta_i}}[X^i_{t+\Delta t}|X^i_{t}]$ for a single realization.  The series appears to come from a stationary distribution (making the autocorrelation functions measured over the entire time series meaningful).  The bottom inset zooms in on the autocorrelation function in order to show that systematic difference between the time series do not seem to ``persist" indicating the the deviations from the diffusion process can be modeled as a single stochastic process.   }
\end{figure}

In regards to multiscale modeling, it is interesting to note that
Figure \ref{fig:ACcombo} makes it appear that the conditional
residuals ($:= X^i_{t+\Delta
t}-\mathbb{E}^{\mathbb{P}_{\theta_i}}[X^i_{t+\Delta t}|X^i_{t}])$
, where the superscripts are meant to distinguish between global
diffusion models and paths associated with different SMD
realizations) come from a single stationary process.  The
empirically measured autocorrelation function of the series above
are all similar and the  differences between the autocorrelation
appear to be random (indicating that a single underlying
autocorrelation function would be a useful approximation of the
conditional residual process).  This suggests that the key
differences in the equilibrium configurations (e.g. different
positions of the deca-alanine atoms) used to start an SMD
simulation determine the coefficient functions of the different
diffusion processes, but the deviations from the collection of
diffusion models  are due to the fast time scale motions
``contaminating" the diffusions in a ``statistically similar"
fashion irrespective of the details of the initial configuration.
To get more realistic surrogate models one could appeal to some
well established time series procedures (ARMA,ARCH, or GARCH
modeling \cite{hamilton}) to model the contamination processes
(which could possibly use data pooled together from different time
series) and then utilize the collection of oversimplified
diffusion models to estimate the conditional expectations yielding
a ``two-scale" type approximation which can possibly be used to
incorporate a more realistic noise into the surrogate model; this
would also be an interesting direction to investigate in the
future.




\section{\label{s_conc} Conclusions}

We have demonstrated a methodology by which one can use local
diffusion modeling in order to construct a global nonlinear SDE
which can be used to approximate the time series that comes out of
a SMD process.  It was also shown that the diffusion approximation
estimated in the deca-alanine example studied appears to come from
a single family of diffusion models (the specific family was
demonstrated to depend  on the sampling frequency used to
calibrate the models).  The statistical validity of the surrogate
models was tested using the PIT and it was shown that the simple
diffusion process appears to be a reasonable (albeit strongly
rejected by large sample hypothesis tests) approximation of the
true process and it was demonstrated that the surrogate models
capture many of the salient features needed to reproduce a PMF.
Some extensions that could utilize the knowledge of researchers
involved with empirical Bayes methods and growth curve analysis
were also discussed.

\section{Acknowledgements}
This work was supported by a Ford Foundation/NRC Fellowship. The
author is grateful for useful discussions and suggestions from
I.G. Kevrekidis and C. Jarzynski.

\bibliography{LLdeca_ii}

\end{document}